\documentclass[fleqn,usenatbib]{mnras}

\usepackage{newtxtext,newtxmath}
\usepackage{comment}
\usepackage[T1]{fontenc}

\DeclareRobustCommand{\VAN}[3]{#2}
\let\VANthebibliography\thebibliography
\def\thebibliography{\DeclareRobustCommand{\VAN}[3]{##3}\VANthebibliography}

\usepackage{graphicx,epstopdf}	
\usepackage{amsmath}	

\title[BFS~10: A nascent bipolar H\,{\normalsize \textit{II}} region]{BFS~10: A nascent bipolar H\,{\Large \textbf{II}} region in a filamentary molecular cloud}

\author[N. Larose \& C. R. Kerton]{
Nicholas Larose$^{1}$\thanks{E-mail: nrlarose@iastate.edu}
and C. R. Kerton$^{1}$
\\
$^{1}$Iowa State University, Department of Physics \& Astronomy, 2323 Osborn Dr., Ames, IA 50011, U.S.A.
}

\date{Accepted XXX. Received YYY; in original form ZZZ}

\pubyear{2022}

\begin{document}
\graphicspath{{./}{figures/}}

\label{firstpage}
\pagerange{\pageref{firstpage}--\pageref{lastpage}}
\maketitle


\begin{abstract}
We present a study of the compact blister \ion{H}{ii} region BFS~10 and its highly filamentary molecular cloud. We utilize $^{12}$CO observations from the Five College Radio Astronomy Observatory to determine the distance, size, mass, and velocity structure of the molecular cloud. Infrared observations obtained from the UKIRT Infrared Deep Sky Survey and the \emph{Spitzer} Infrared Array Camera, as well as radio continuum observations from the Canadian Galactic Plane Survey, are used to extract information about the central \ion{H}{ii} region. This includes properties such as the ionizing photon rate and infrared luminosity, as well as identifying a rich embedded star cluster associated with the central O9~V star. Time-scales regarding the expansion rate of the \ion{H}{ii} region and lifetime of the ionizing star reveal a high likelihood that BFS~10 will develop into a bipolar \ion{H}{ii} region. Although the region is expected to become bipolar, we conclude from the cloud's velocity structure that there is no evidence to support the idea that star formation at the location of BFS~10 was triggered by two colliding clouds. A search for embedded young stellar objects (YSOs) within the molecular cloud was performed. Two distinct regions of YSOs were identified; one region associated with the rich embedded cluster and another sparse group associated with an intermediate mass YSO.  
\end{abstract}

\begin{keywords}
\ion{H}{ii} regions -- ISM: clouds -- stars: pre-main-sequence -- infrared: stars
\end{keywords}


\section{Introduction} \label{sec:intro}

This paper presents a study of the compact Galactic \ion{H}{ii} region BFS~10 \citep*{bfs82}. The region hosts a rich embedded massive star cluster, and is currently a blister \ion{H}{ii} region that will evolve into a bipolar region on a very short time-scale. 

Bipolar \ion{H}{ii} regions are \ion{H}{ii} regions that are density bounded in two opposite directions, and thus, when viewed at wavelengths tracing the distribution of ionized gas, can exhibit a bipolar shape. This is in contrast to the common blister morphology that occurs when the \ion{H}{ii} region is density bounded in only one direction \citep{isr78,tt82}. While a bipolar morphology will arise naturally for a region evolving in a filamentary or sheet-like molecular cloud (e.g., see the simulations by \citealt*{bty79}), it has also been suggested that a bipolar \ion{H}{ii} region morphology is a potential signature of massive star formation caused by colliding molecular clouds \citep{whit18}.

Once a massive OB star forms and starts to ionize the surrounding interstellar medium, the resulting \ion{H}{ii} region evolves through the ultra compact, compact, and evolved stages, with typical size scales of $0.1$, $1$, and $\gg 1$~pc respectively \citep{hab79, chu02}. The compact stage is interesting as it affords us the first good look at the stellar content of the \ion{H}{ii} region, which is typically highly obscured at the ultra compact stage.  In addition, unlike the evolved stage, the surrounding, parsec-scale, molecular material has not yet been disrupted, so the large-scale molecular environment that led to massive star formation can be examined. 

In \autoref{sec:obsdata} we briefly describe our molecular line observations as well as the various archival data sets used. Next, in \autoref{sec:analysis}, we explore the physical properties of the \ion{H}{ii} region, the surrounding filamentary molecular cloud, and the rich embedded star cluster. In particular, we explore whether or not there is any evidence suggesting that a molecular cloud collision is ongoing. In \autoref{sec:discuss} we compare the molecular cloud structure with other molecular clouds, and use a simple model to demonstrate that the three-dimensional shape of the molecular cloud is filamentary as opposed to sheet-like. We then show how BFS~10 will develop a bipolar morphology on a very short time-scale, and how the entire molecular cloud will be dispersed over the lifetime of the O star powering the region. We also examine the young stellar object population found within the embedded cluster and throughout the molecular cloud. Finally, in \autoref{sec:conclusions} we present our conclusions.

\section{Observations \& Archival Data} \label{sec:obsdata}

Molecular line, $^{12}$CO $(J=1-0)$ and $^{13}$CO $(J=1-0)$, observations of a $0.75 \times 0.5$ degree area, including the BFS 10 location, were obtained using the Five College Radio Astronomy Observatory (FCRAO) 14-m telescope in spring 2003. Details of the observing setup and data reduction are given in \cite*{ker04}. The reduced data cubes used in this study have velocity coverage of $V_\mathrm{LSR} \sim -110$ to $\sim +20$ km~s$^{-1}$ with a channel spacing of 0.13 km~s$^{-1}$. The spatial resolution (beam FWHM) is 45 arcsec, and the velocity resolution is 1 km~s$^{-1}$. The sensitivity per channel (1$\sigma$, $T_\mathrm{mb}$ scale) is 0.3~K ($^{12}$CO) and 0.15~K ($^{13}$CO).

In addition to our molecular line observations, we gathered a large amount of archival data on BFS~10. Radio continuum images at 1420 and 408 MHz were obtained from the Canadian Galactic Plane Survey \citep[CGPS;][]{tay03}. These images, which include short-spacing data, have a spatial resolution of $\sim 1$ and 3.5 arcmin, and a noise level of $\sim 0.3$ and $3$ mJy beam$^{-1}$, at 1420 and 408 MHz respectively. All CGPS data are available via the Canadian Astronomy Data Centre (CADC).

At mid- and far-infrared wavelengths, the Mid-Infrared Galaxy Atlas \citep[MIGA;][]{ker00} and the \emph{IRAS} Galaxy Atlas \citep[IGA;][]{Cao97}, provided $0.5 - 1$ arcmin resolution images at 12, 25, 60 and 100~$\mu$m. These data were also obtained via the CADC. Higher resolution (6 arcsec) 24~$\mu$m images from the \emph{Spitzer} Mapping of the Outer Galaxy \citep[SMOG;][]{smog08} survey were obtained from the NASA/IPAC Infrared Science Archive (IRSA). The rms noise level of the data in the area around BFS~10 is 5.2 MJy~sr$^{-1}$. \emph{Herschel} PACS images of the BFS 10 region at 70 and 160 $\mu$m were obtained using the \emph{Herschel} High-Level Images (HHLI) interface at IRSA. In the area around BFS~10, these images had 10 and 14 arcsec spatial resolution (FWHM) at 70 and 160 $\mu$m respectively, and a 0.01 Jy pixel$^{-1}$ rms noise level in both bands.

Finally, in the near-infrared, we used point source catalogues and images at 3.6, 4.5 and 5.8 $\mu$m from the \emph{Spitzer} SMOG and  GLIMPSE360  \citep{whi2011} surveys, and in the $J$, $H$, and $K$ bands from the United Kingdom Infrared Deep Sky Survey (UKIDSS) Galactic Plane Survey \citep[GPS;][]{lucas2008}. GLIMPSE360 data were obtained from IRSA, and UKIDSS-GPS data were obtained from the Wide Field Camera Science Archive \citep{ham08}. 

\section{Analysis} \label{sec:analysis}

\subsection{Compact \ion{H}{ii} Region} \label{sec:HII}

\subsubsection{Distance}
\begin{figure}
    \centering
    \hspace*{-0.9cm}
    \includegraphics[width=1.2\linewidth]{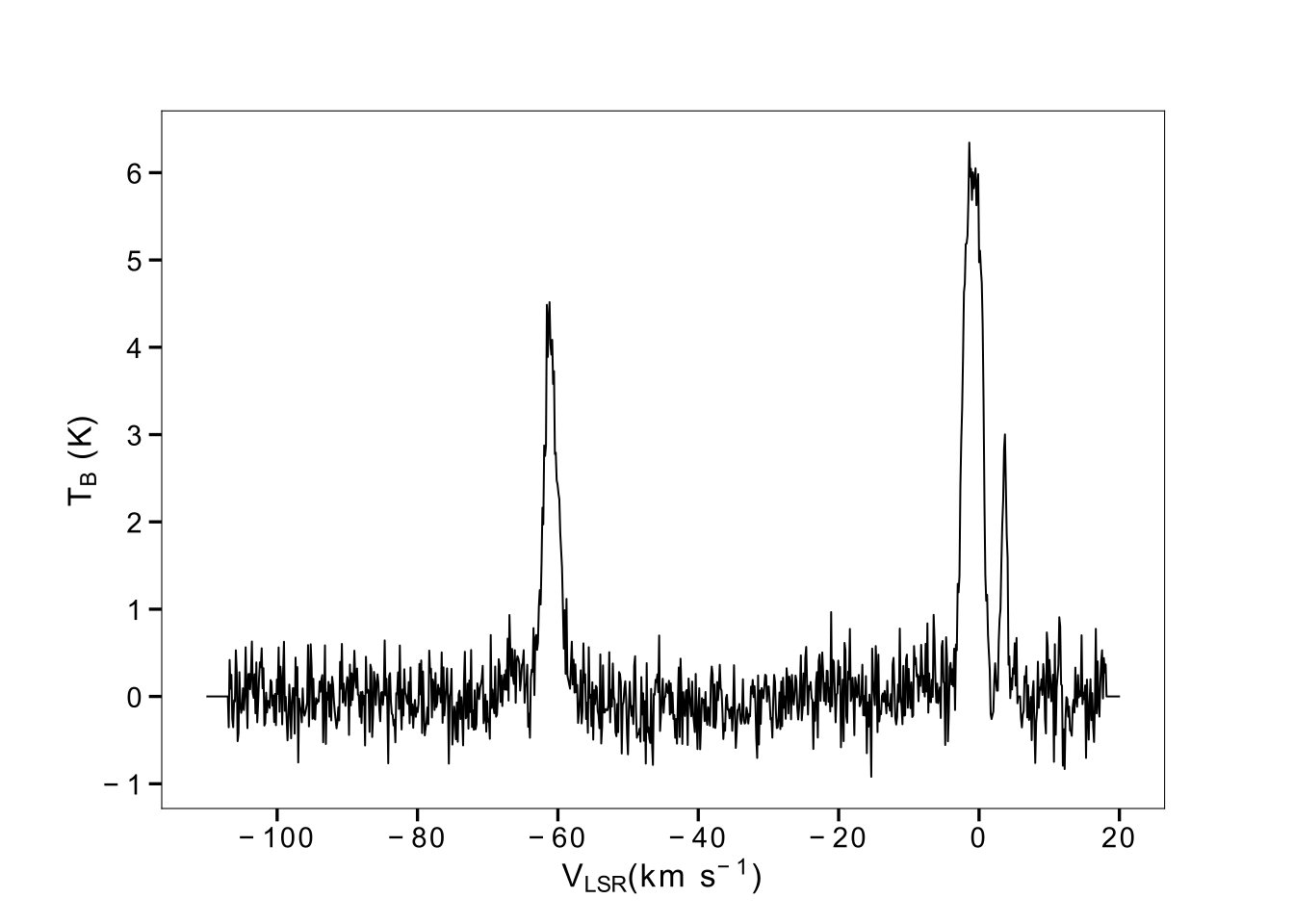}
    \vspace*{-0.5cm}
    \caption{$^{12}$CO average spectrum over a 2.62 x 2.62 arcmin region surrounding BFS 10. Emission associated with the \ion{H}{II} region appears as the strong peak located at $-61$~km~s$^{-1}$. The stronger peaks located near 0~km~s$^{-1}$ are emission from unrelated local molecular clouds.}
    \label{fig:Spectrum}
\end{figure}

The $^{12}$CO spectrum (see \autoref{fig:Spectrum}) towards BFS~10 shows a single strong peak at $V_{\mathrm{LSR}} = -61$~km~s$^{-1}$. To find a distance to BFS~10 we used the Revised Kinematic Distance Tool \citep[RKDT; ][]{Reid}, which combines a kinematic model, a spiral arm model based on maser parallax distances, and positional information to derive a probabilistic estimate of the distance. For BFS~10, the RKTD returns $5.99\pm0.71$~kpc as the best estimate of the distance (94 per cent probability). This is in agreement with the \citet{Russeil} distance of $6.39\pm0.67$~kpc and the \citet{fos2015} distance of $6.18\pm1.24$~kpc, which are both based on the spectroscopic parallax of the exciting star of the \ion{H}{ii} region.

\subsubsection{Size and Radio Spectral Index}
BFS~10 is a slightly resolved compact source in the 1420 and 408~MHz CGPS images. We used the Dominion Radio Astrophysical Observatory (DRAO) Export Software Package program {\sc fluxfit} \citep{hig97} to obtain size and flux density measurements that incorporated information about the beam size and orientation. For a size estimate we used the lower noise, higher-resolution 1420~MHz image and measured an angular diameter (FWHM) of $1.301\pm0.003 \times 1.208\pm0.003$ arcmin. Correcting for the beam size results in an angular diameter (FWHM) of $0.895\pm0.004 \times 0.860\pm0.005$ arcmin. This corresponds to a physical diameter of $1.56\pm0.18 \times 1.49\pm0.18$~pc, using a distance of $5.99\pm0.71$~kpc, or an average physical radius (defined as $r=2\sigma\sim0.849$~FWHM) of $1.3\pm0.16$ pc.

The flux density at 1420 and 408~MHz is $F_\mathrm{1420} = 203\pm8$~mJy and $F_\mathrm{408}=182\pm10$~mJy respectively. The 408--1420~MHz radio spectral index, defined as $F_\nu \propto \nu^\alpha$, is $\alpha = +0.09\pm0.03$, which is consistent with optically thin, thermal radio emission.

\begin{figure}
    \centering
    \vspace*{-0.6cm}
    \hspace*{-0.25cm}
    \includegraphics[width=1.125\linewidth]{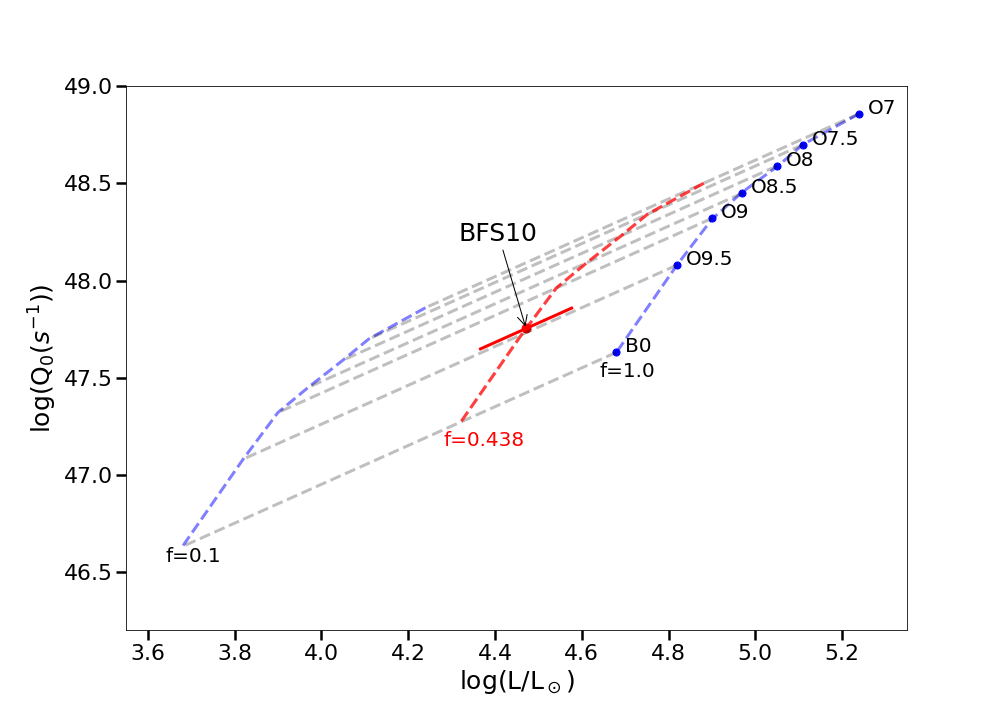}
        \vspace*{\fill}
    \includegraphics[width=1\linewidth]{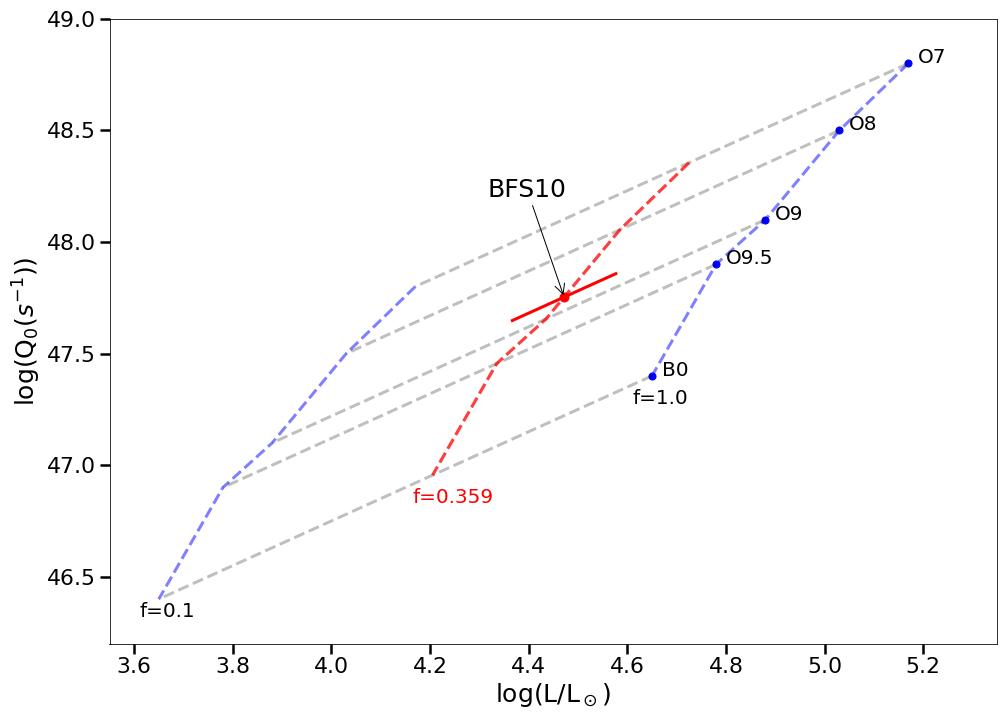}
    \vspace*{-0.15cm}
    \caption{Ionizing photon rate ($Q_0$) vs. bolometric luminosity ($L/L_{\odot}$) for OB-star atmospheric models from \citet{Panagia} (top) and \citet{Crowther} (bottom). Dashed blue lines show values expected for different spectral types given geometric covering factors of $f=0.1$ and $f=1$. A linear interpolation is used between tabulated values for the indicated main-sequence spectral types. Dashed black lines represent the locus followed by the models as $f$ is varied. The best-fit covering factor is displayed as a red dashed line passing through the observed $Q_0$ and $L/L_{\odot}$ data point for BFS~10.
    }
    \label{fig:qlplot}
\end{figure}

\subsubsection{Ionizing Photon Rate and Infrared Luminosity}\label{sec:IPRIRL}

The radio continuum flux at 1420~MHz can be used to derive the ionizing photon rate ($Q_0$;  photons s$^{-1}$) of the star(s) powering the \ion{H}{ii} region. Following \cite{Mat76}, we have:
\begin{equation}
     \left(\frac{Q_0}{\textrm{s}^{-1}}\right) = 7\times10^{46}\left(\frac{S}{\mathrm{Jy}}\right)\left(\frac{d}{\mathrm{kpc}}\right)^2\left(\frac{\nu}{\mathrm{GHz}}\right)^{0.1}\left(\frac{T_e}{10^4\,\mathrm{K}}\right)^{-0.45},
    \label{eqn:rate}
\end{equation}
where $S$ is the flux density in Jy, $d$ is the distance in kpc, $\nu$ is the frequency in GHz, and $T_e$ is the electron temperature in units of 10$^4$~K. Adopting $S=0.203$, $d=5.99$, $\nu = 1.4$, and $T_e=1$ we find $\log\left(Q_0\right) = 47.75\pm0.10$.

The infrared luminosity ($L_{\mathrm IR}$) of an embedded \ion{H}{ii} region is a good estimate for the luminosity of the exciting star(s). We performed photometry on BFS~10 from 12 to 160~$\mu$m. The resulting spectral energy distribution (SED) was then extrapolated from 160 to 1000~$\mu$m using a single temperature greybody fit, $F_\nu \propto \nu^\beta B_\nu(T_d)$, where $B_\nu(T_d)$ is the Planck function evaluated at dust temperature $T_d$,  and the $\nu^\beta$ term accounts for the emissivity of the dust grains. The value of $\beta$ depends on the exact composition (silicate, carbonaceous, composite) and structure (crystalline, amorphous, layered) of the grains, and is known from both theory and observation to vary between 1 and 2 \citep{leq05, whi03, den98}. For this study we use $\beta=1.5$, although we note that the luminosity we derive is not strongly dependent on the value used: the luminosity is identical for $\beta = 1.3 - 2.0$, and is only 0.02 dex lower for $\beta = 1.0$.  The ratio of $\nu^{1.5} B_\nu(T_d)$, evaluated at 100 and 160~$\mu$m, was matched to the observed $F_{100}/F_{160}$ flux density ratio using a dust temperature of $T_d = 27.02$~K. The greybody curve ($F_\nu \propto \nu^{1.5} B_\nu(27.02~\mathrm{K})$) was scaled to match the $F_{160}$ data point, and was evaluated at 250, 500 and 1000 $\mu$m. Using a distance of $5.99\pm0.71$ kpc, a numerical integration of the SED, including a $+0.1$~dex correction factor from $L_{\mathrm IR}$ to the bolometric luminosity ($L_\mathrm{bol}$) as outlined in \cite*{ker99},  resulted in $\log\left(L_\mathrm{bol}/L_\odot\right) = 4.47\pm0.11$.

\subsubsection{Spectral Type and Covering Factor} \label{sec:CoveringFactor}

\begin{figure}
    \hspace*{-1.45cm}
    \includegraphics[width=1.4\columnwidth]{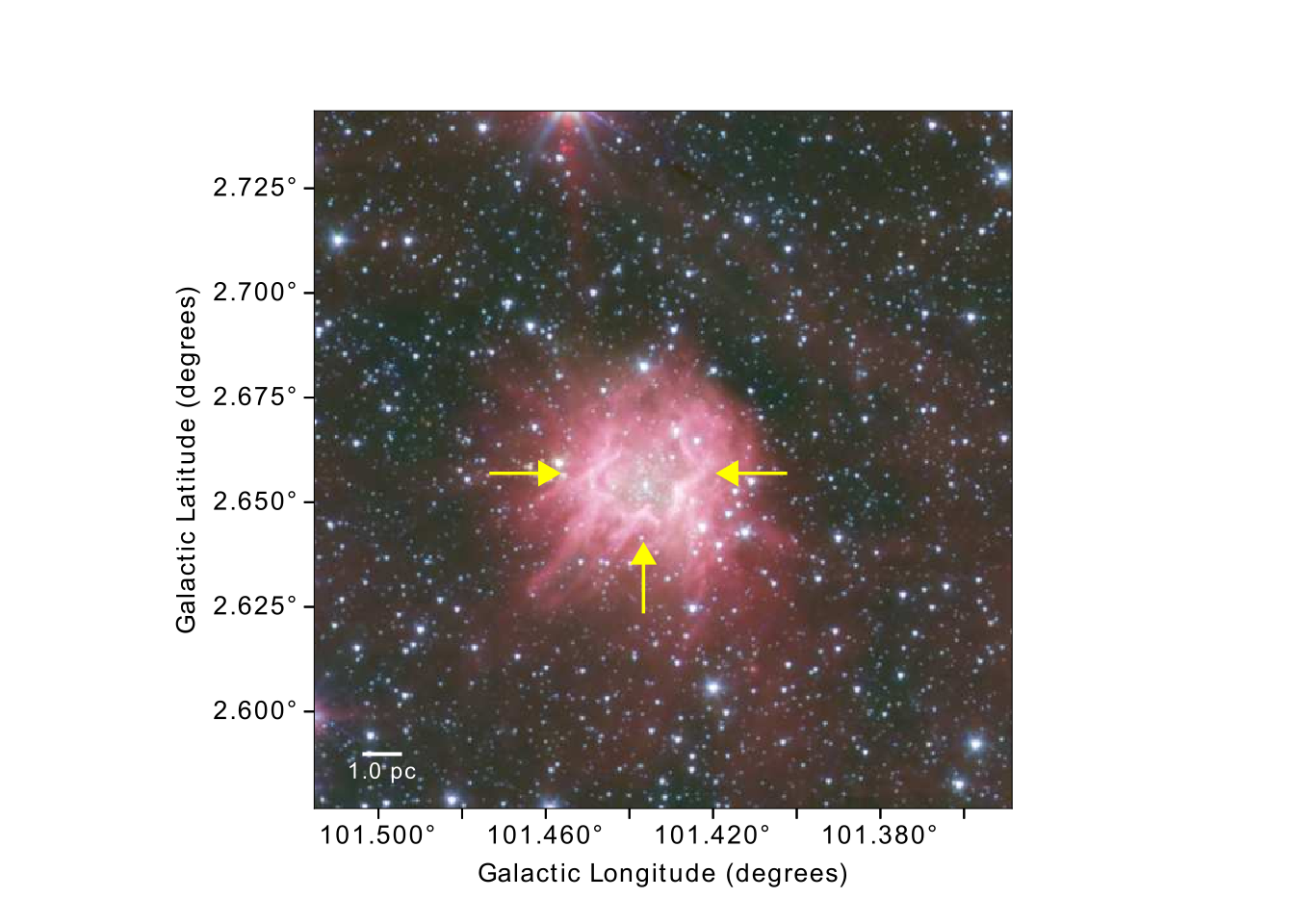}
    \vspace*{-0.75cm}
    \caption{\emph{Spitzer} IRAC composite image of BFS~10. Yellow arrows indicate the location of the ionization-bounded rims tracing where the \ion{H}{ii} region is interacting with the surrounding molecular cloud. Note the lack of bright rims at the top of region showing the \ion{H}{ii} has blown out in that direction. The central star cluster is easily identified at these near-infrared wavelengths, which implies this region is not heavily embedded along the line of sight. The IRAC 3.6, 4.5 and 5.6 $\mu$m images are mapped to Blue, Green and Red channels respectively. 
    }
    \label{fig:bfs10ir}
\end{figure}

The observed ionizing photon rate and bolometric luminosity are related to values derived from stellar atmosphere models by:
\begin{equation}
    \log(Q_{0,\mathrm{obs}}) = \log(Q_{0,\mathrm{model}}) + \log(f)
    \label{eqn:qf}
\end{equation}
and
\begin{equation}
    \log(L_{\mathrm{bol},\mathrm{obs}}) = \log(L_{\mathrm{bol},\mathrm{model}}) + \log(f),
    \label{eqn:lirf}
\end{equation}
where $f$ is a factor accounting for expected deviations from the model values. In this study we assume $f$ represents the geometrical covering factor of the surrounding material (e.g., a deeply embedded, ionization bounded, \ion{H}{ii} region would have $f\sim1$), and that $f$ is the same for both the radio and infrared observations. Using the atmospheric models from \cite{Crowther}  we find, using \autoref{eqn:qf} and \autoref{eqn:lirf}, that the infrared and radio observations are consistent with an O8--O9~V star with $f = 0.36^{+0.08}_{-0.05}$ (see \autoref{fig:qlplot}). For comparison, we repeated the analysis with models from \cite{Panagia} and found the observations are consistent with an O9--O9.5 V star with $f=0.44^{+0.09}_{-0.07}$. In both cases the error estimate is dominated by the uncertainty in the distance. \citet*{Russeil} used a medium resolution optical spectrum to classify the central star of BFS~10 as O9~V, consistent with the range of potential spectral types suggested from our radio and infrared analysis. The low value of $f$ we derive, regardless of the atmospheric model used, is appropriate for a partially embedded \ion{H}{ii} region. The appearance of BFS~10 in the near infrared (see \autoref{fig:bfs10ir}) suggests this is a likely morphology for the region; if the region were more heavily embedded, the central cluster would not be as visible. Additionally, bright ionization-bounded rims can be seen surrounding the \ion{H}{II} region, except for the upper part where it has blown out of the molecular cloud.

\subsection{Filamentary Molecular Cloud} \label{sec:mol}

\subsubsection{Size, Mass, and Velocity Structure}

\begin{figure}
\centering
\hspace*{-0.5cm}   
\includegraphics[width=1.2\columnwidth]{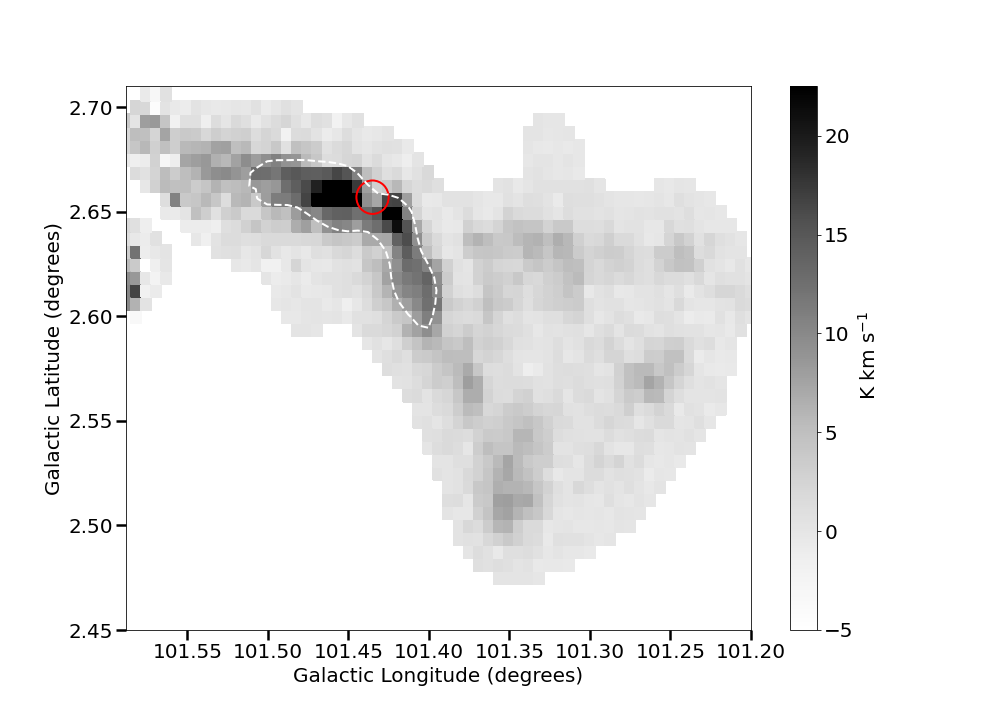}

\hspace*{-4.85cm}
\vspace*{-0.5cm}
\includegraphics[width=1.7\columnwidth]{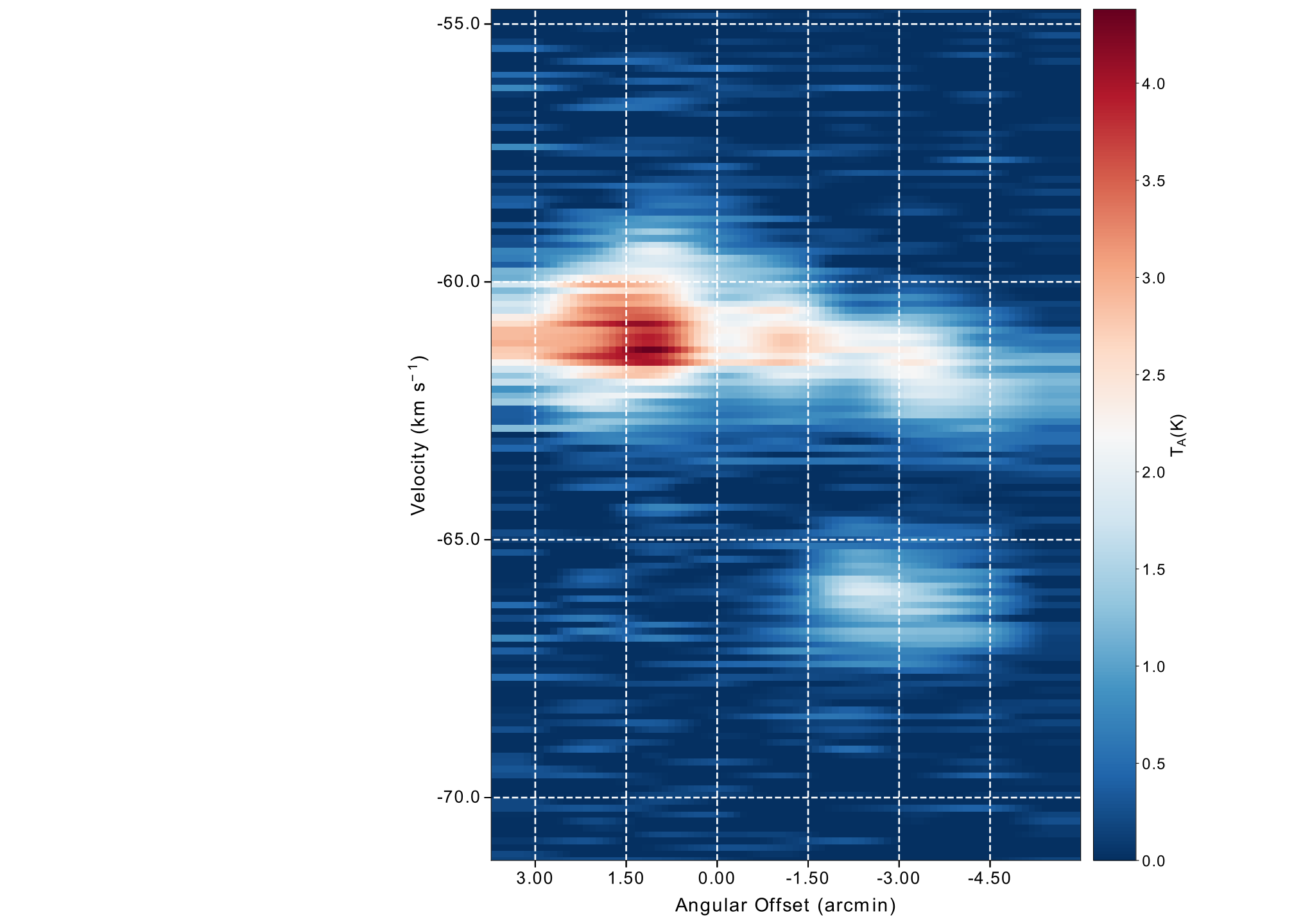}
\caption{(Top) $^{12}$CO zeroth moment (clipped integrated intensity) map. The white contour at 1.67~K~km~s$^{-1}$ outlines the general shape of the filamentary cloud. A local minimum in the zeroth moment map, associated with the position of BFS~10, can be seen near the bend in the cloud. The red circle indicates the location and approximate size of the \ion{H}{II} region. (Bottom) Position-velocity diagram of the molecular cloud, constructed using nine box-averaged spectra evenly spaced along the molecular cloud centerline. A linear interpolation has been applied for clarity. The angular offset of zero arcmin marks the location of the \ion{H}{II} region, and positive offset is in the direction of increasing Galactic longitude. There are no features in the p-v diagram suggesting that a cloud-cloud collision has occurred (see text for details). The bright emission (left of the \ion{H}{II} region) shows the location of the secondary region of star formation, with weaker emission (right of the \ion{H}{II} region) coinciding with the lack of star formation in the lower leg.}
    \label{fig:MomMap}
\vspace*{-0.1cm}
\end{figure}

We constructed moment maps of our $^{12}$CO data cube following the optimized moment masking technique described in \citet{Dame}. Integration in velocity space was between $-58.9$ and $-62.8$ km s$^{-1}$. The resulting zeroth and first moment maps, in the $23.3 \times 15.6$ arcmin region around BFS~10 are shown in \autoref{fig:MomMap}.

The zeroth moment map shows that the molecular cloud associated with BFS~10 has an elongated, dog-leg morphology. The \ion{H}{ii} region is located at the bend in the molecular cloud and above the centre line of the integrated emission from the cloud.

The length of the cloud, as measured along the centre line, is 9.4 arcmin (16.3 pc), and the average width of the cloud is 1.4 arcmin (2.5 pc). Assuming a cylindrical, or filamentary, morphology we estimate the volume to be $80\pm19$~pc$^{3}$. A useful metric for describing elongated structures is an effective size ($s_{\mathrm{eff}}$):
\begin{equation}
    s_{\mathrm{eff}} = 2\left(\frac{A}{\pi}\right)^{\frac{1}{2}},
    \label{eqn:size}
\end{equation}
where $A$ is the area of the cloud in the zeroth moment map defined by counting contiguous pixels above a threshold of $\sim 3$~K~km~s$^{-1}$. For the BFS~10 cloud we find $s_{\mathrm{eff}}=7.2$ pc.

The $^{12}$CO to $^{13}$CO brightness temperature ratio for the BFS~10 molecular cloud is only $\sim 3.5$, which shows that the $^{12}$CO emission is optically thick. The $^{12}$C to $^{13}$C abundance ratio, which in the optically thin case reflects the  brightness temperature (or equivalently the column density) ratio, is an order of magnitude larger \citep{mm2015,mil2005}. To estimate the mass ($M$) of the molecular cloud we converted our zeroth moment map to a column density map using the standard Galactic CO-to-H$_2$ conversion factor $X_{\mathrm{CO}} = 2\pm0.6\times$10$^{20}$~(K~km~s$^{-1}$)$^{-1}$ \citep*{bol2013,szh2016}.

$X_{\mathrm{CO}}$ is known to increase with decreasing metallicity, with a very strong increase seen for metalicities below 0.5 Z$_\odot$ (see, e.g., Figure 9 in \citealt{bol2013}). BFS~10 has a Galactocentric distance ($R_G$) of 10.9 kpc (for R$_\odot = 8.0$~kpc). The metallicity gradient in the Milky Way is approximately $-0.05$~dex kpc$^{-1}$ \citep{bal2011} meaning that the metallicity at the Galactocentric distance of BFS~10 is still close to solar, $\sim 0.86$~Z$_\odot$. This can be contrasted with far-outer Galaxy \ion{H}{ii} regions like WB89~361 ($R_G= 18$~kpc) and WB89~529 ($R_G = 19$~kpc) \citep{rud1996}, where the metallicity would be $\lesssim 0.5$ Z$_\odot$ and some correction for metallicity would perhaps be appropriate.

Using a mean mass per H$_{\mathrm 2}$ molecule, $\mu_{\mathrm{H2}}=2.8$, and a distance of 5.99 kpc, spatial integration of the column density map results in $M=3.7\pm1.4 \times 10^{3}$ M$_{\odot}$. Given the volume derived above we find an average density of n$_{\mathrm H2} = 6.7\pm3.5\times10^2$~cm$^{-3}$. 

As a check on the $^{12}$CO derived mass we followed the techniques described in \citet{mar2019} and \citet{ker04} to derive a mass estimate from the $^{13}$CO data. The $^{13}$CO zeroth moment map looks essentially like the $^{12}$CO map shown in \autoref{fig:MomMap}, but the total extent of the cloud (especially the lower latitude part of the dog-leg structure) is reduced as fainter emission is lost in the noise (the per channel noise level is lower by about a factor of two compared to the $^{12}$CO emission, but the intensity ratio is lower by a factor of $\sim 3$ -- 4). The $^{13}$CO column density is related to the integrated $^{13}$CO intensity (in the zeroth moment map) by:
\begin{equation}
N(\mathrm{^{13}CO}) = 7.3\times10^{14} \int T_{MB}\, dv \, \mathrm{cm}^{-2}.
\end{equation}
Using $R_{12}/R_{13} = 86$ \citep{mil2005} and H$_2/^{12}$CO $= 1.25\times10^{4}$ \citep{bla1987} we find $M = 2.0\pm0.5\times 10^{3}$~M$_\odot$, where the quoted uncertainty reflects only the distance uncertainty.

We can also compare the column-density derived estimates with the virial mass estimate given by:
\begin{equation}
M_\mathrm{vir} = 1040 \times R \times \sigma^2 ,     
\end{equation}
where $R$ is the cloud radius (or similar size scale) in parsecs, and $\sigma$ is the velocity dispersion in km~s$^{-1}$ \citep{szh2016}. The numerical factor depends on the density structure of the cloud, but it is always of order $10^3$. Using $R=3.5$ and $\sigma = 0.85$, we find $M_\mathrm{vir} = 2.9\pm0.7\times 10^3$~M$_\odot$ (the quoted uncertainty is again just from the distance uncertainty). We conclude that the $X_\mathrm{CO}$ mass estimate is a reasonable description of the cloud mass given its  general agreement with both the $^{13}$CO and the virial mass estimate.

Given its clear association with massive star formation, it is interesting to note that the BFS~10 molecular cloud has properties similar to an average infrared dark cloud (IRDC): $\overline{M}\sim10^3$ M$_\odot$, $\overline{n}_\mathrm{H2} \sim 10^{3}$ cm$^{-3}$, and $\overline{s}_\mathrm{eff} \sim 5$ pc \citep{Simon06}. The BFS~10 molecular cloud is not visible as an IRDC due to the lack of a strong mid-infrared background given its location in the outer Galaxy. Molecular clouds with the size and mass of the BFS~10 molecular cloud are common in the outer Galaxy \citep{Heyer98}; however, as we discuss in \autoref{sec:discuss}, the highly elongated nature of the cloud is not commonly seen in outer Galaxy molecular clouds.

A position-velocity (p-v) diagram was used to examine the large-scale velocity structure of the cloud. The p-v diagram, presented in the lower panel of \autoref{fig:MomMap}, was constructed by averaging the spectrum in nine, 1.25 arcmin square boxes evenly spaced at 1.25 arcmin intervals along the centerline of the cloud.The \ion{H}{II} region is located at an angular offset of 0, and positive offset is in the direction of increasing Galactic longitude. The molecular cloud hosting BFS~10 is the structure centered at $-61$~km s$^{-1}$, and at the position of the \ion{H}{ii} region there is only a single velocity component (see also \autoref{fig:Spectrum}). At lower longitudes an additional velocity component, located at $-66$~km~s$^{-1}$, is visible. Inspection of the p-v diagram reveals the lack of a broad "bridge" structure connecting these two velocity components, suggesting that they arise in two separate clouds, and that they are not associated with a cloud-cloud collision \citep{Haworth2015}.

We were also interested in determining if the two sections of the molecular cloud, on either side of the \ion{H}{ii} region, had distinctly different velocities analogous to the \citet{Loren76} observations of NGC~1333. In the case of NGC~1333, the two velocity components are interpreted as two partially colliding molecular clouds, with star formation occurring in the collision region. In our case, the p-v diagram shows there is no spatially offset red- and blue-shifted emission;  the difference in the average intensity weighted velocity (on either side of the \ion{H}{II} region) is only $0.6$~km s$^{-1}$, while the average line-width across the cloud is $2.3$~km s$^{-1}$. 

The lack of a broad bridge structure in our p-v diagram, the single velocity component along the line of site to the \ion{H}{ii} region, and the lack of offset red/blue shifted emission, all suggest that the observed high mass star formation in BFS~10 was not due to a cloud-cloud collision.

\subsection{Embedded Star Cluster} \label{sec:cluster}

\begin{figure*}
    \centering
    \begin{minipage}{0.7\textwidth}
      \centering
      \hspace*{-2.9cm} 
      \includegraphics[width=\linewidth]{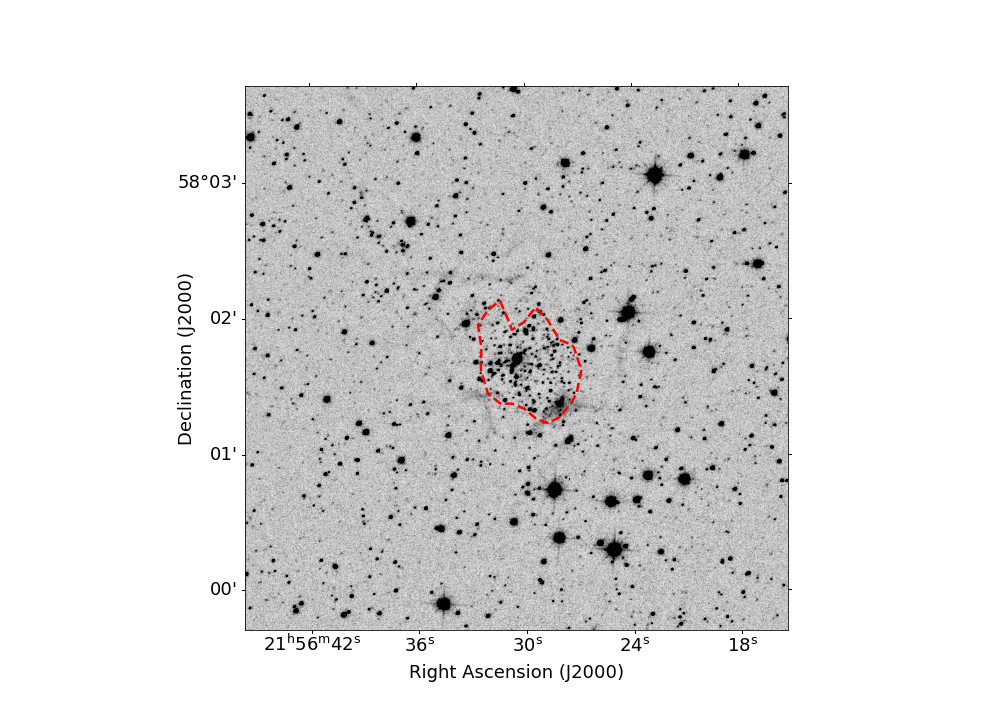}
      \label{fig:test1}
    \end{minipage}%
    \begin{minipage}{.6\textwidth}
      \centering
      \hspace*{-9cm} 
      \includegraphics[width=\linewidth]{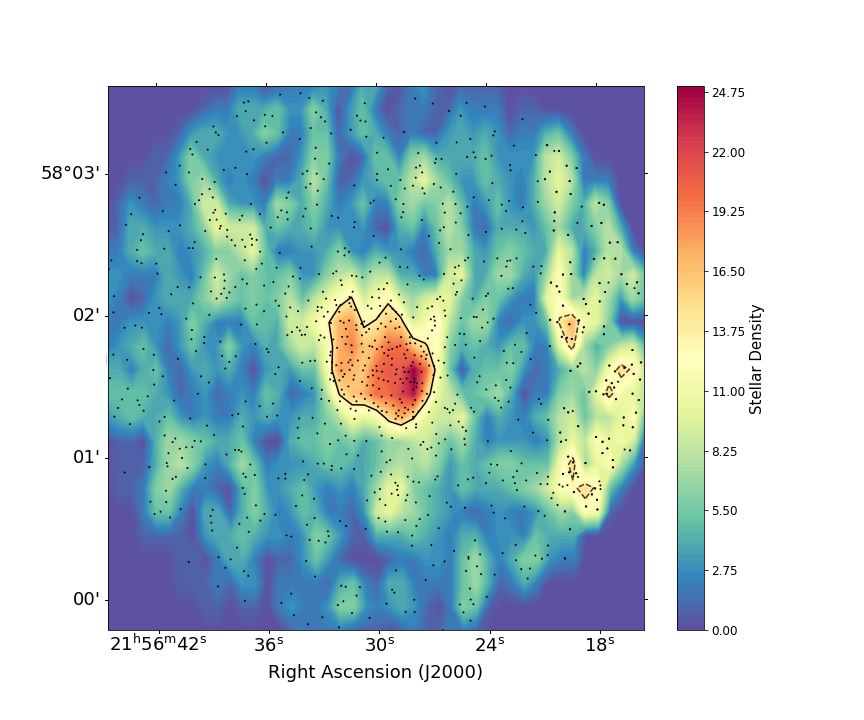}
      \label{fig:test2}
    \end{minipage}
    \vspace*{-0.6cm}
    \caption{(Left) UKIDSS K-band image of the rich embedded star cluster associated with BFS~10. The cluster extent, defined by the cluster-finding algorithm is shown by the red dashed contour. Bright rims associated with the molecular cloud -- \ion{H}{II} region interface can be seen in the image. (Right) Stellar surface density density map of 4 arcmin cluster search region, measured in units of stars per $20\times20$ arcsec search area. Dashed contours represent cluster candidates with 3$\sigma$ enhancements to the stellar background. The solid black contour, which defines the extent of the embedded star cluster, represents the only cluster candidate with a 5$\sigma$ enhancement to the stellar background.
    }
\label{fig:cluster}
\end{figure*}

A compact star cluster surrounding the exciting star of BFS~10 is clearly visible in UKIDSS K-band images of the region (see \autoref{fig:cluster}). To determine the cluster properties we applied a cluster identification method outlined by \cite*{Carpenter}. A $4\times4$~arcmin K-band image was first subdivided into $20\times20$~arcsec bins. Cluster candidates were determined by comparing the background stellar surface density to the stellar surface density in each bin, where the background stellar surface density was first determined by fitting a Poisson distribution to the lower density bins. A cluster was identified if the total number of stars within 3$\sigma$ contours represented a 5$\sigma$ enhancement to the stellar background. Using this method we estimate that the cluster contains 151$\pm$8 stars within a 0.92$\pm$0.09 pc cluster radius. 

As expected, given the presence of the O star, this is a very rich cluster. For context, only two of the 19 embedded clusters associated with the W3/4/5 \ion{H}{ii} regions identified by \citet{Carpenter}, contain more stars. We note that this compact cluster corresponds to one of the subclusters of cluster \#60 identified in the \citet{win2019} study of the SMOG region. 

\section{Discussion} \label{sec:discuss}

\subsection{The Filamentary BFS 10 Molecular Cloud} \label{sec:cloudcontext}

The molecular cloud containing BFS~10 is highly elongated with an aspect ratio (AR) of only 0.15. For context, the average AR of the $^{12}$CO molecular clouds in the FCRAO Outer Galaxy Survey \citep[OGS;][]{hey1998} is 0.58 with a standard deviation of 0.18 \citep*{bkp03}. Of the 13100 OGS molecular clouds identified by \citet{bkp03} only 20 (0.15 per cent) have AR$\le0.15$.

Defining $\hat{x}$ as the direction along the long axis of the molecular cloud, $\hat{y}$ as the direction to the observer, and $\hat{z}$ as the direction along the short axis of the molecular cloud, we can construct a simple model of the molecular cloud as an elongated slab with a square, $2.9\hat{y} \times 2.9\hat{z}$ pc cross section. The \ion{H}{ii} region itself is modeled as a $2.6\hat{x} \times 2.9\hat{y}  \times 2.4\hat{z}$ pc cavity, and the exciting star is centred in the cavity 1.3 pc from the bottom. This results in the \ion{H}{ii} region being ionization bounded in the $\pm\hat{x}$ and $-\hat{z}$ directions, which corresponds to the directions where we see bright rims (see \autoref{fig:bfs10ir}).  The \ion{H}{ii} region is density bounded in the other three directions, which matches the observed morphology and the fact the embedded star cluster is clearly visible. We find that the geometric covering factor for this model \ion{H}{ii} region is $f=0.39$, which is consistent with the covering factors derived from stellar atmospheric models and observations at radio and infrared wavelengths (see \autoref{sec:CoveringFactor}). This demonstrates that the molecular cloud truly has a three-dimensional filamentary morphology rather than being a sheet-like structure that is being observed edge-on. 
%

\subsection{Evolution to a Bipolar \ion{H}{ii} Region and Beyond}
\label{sec:evolution}

Currently BFS~10 has a classic blister morphology in the plane of the sky. To determine whether or not BFS~10 will ever develop a bipolar morphology (i.e., ionization bounded in two directions, and density bounded in two other directions), we need to estimate the time it will take for the region to blow out (i.e. become density bounded) in the $-\hat{z}$ direction and compare this to the remaining lifetime of the star.  

To estimate the remaining lifetime of the star we used the high-mass star evolutionary models of \citet{Schaerer}. For $\log\left(T_\mathrm{eff}\right) = 4.519$ \citep[O9~V from ][]{Crowther} and $\log\left(L_\mathrm{bol}/L_\odot\right) = 4.47\pm0.11$ we find a best match with their 20~M$_\odot$ model with a current age of 3.65 Myr. The main sequence lifetime for this model is $\approx 7.8$~Myr, so we adopt 4 Myr as an estimate for the remaining lifetime of the star.

The evolution of the ionization bounded side of a blister \ion{H}{ii} region can be described by:
\begin{equation}
\frac{t}{\mathrm{(Myr)}} = 0.04\frac{r_o}{\mathrm{(pc)}}\left[ \left(\frac{r(t)}{r_o} \right)^{5/2} - 1 \right] ,
   \label{eqn:OG}
\end{equation}
where $r_o$ is the radius at $t=0$ and $r(t)$ is the radius at time $t$ \citep*{fra1994}. For BFS~10 we set $r_o = 1.3$~pc and $r(t)=1.8$~pc. Applying \autoref{eqn:OG} we find the time for the region to become bipolar is  $t= 0.08$~Myr. This is only 2 per cent of the star's remaining lifetime, so clearly BFS~10 will rapidly develop a bipolar morphology.

After forming a bipolar nebula, the exciting star of BFS~10 will continue to ionize and compress the two remaining parts of the molecular cloud, which we can roughly model as two constant density (n$_\mathrm{H2} = 670$~cm$^{-3}$) cylindrical filaments ($r\sim1.25$~pc, $l \sim 8$~pc) each located a distance $R\sim 1.8$~pc from the O star.

\citet{B89} (B89 hereafter) developed an analytic model of the interaction of a molecular cloud with an incident ionizing radiation field in terms of the molecular cloud column density and the incident radiation field strength, which are described by the dimensionless parameters $\eta$ and $\Gamma$ respectively (Equations 2.1 and 2.4 in B89). For the BFS~10 filament we find $\log\eta = 2.77$ and $\log\Gamma = -3.69$, which places the clouds in a regime where they will be compressed by the passage of an ionization-shock front (ISF; see Figure 1 in B89). This is in contrast to cases where the clouds would be essentially instantaneously fully ionized (low column density and strong radiation field) or cases where they would be essentially unaffected (high column density and weak incident radiation field). While the B89 models were developed for spherical clouds they note that for clouds elongated along the symmetry axis the ISF propagation is essentially unchanged except for the duration. Applying Equation 2.5 of B89 we find the ISF velocity through the cloud is 8.6 km~s$^{-1}$ (using 11.4~km~s$^{-1}$ as the isothermal sound speed in the ionized gas). This means the filamentary clouds will be compressed and partially ionized in only $\sim 1$~Myr.

\citet{wp21} (WP21 hereafter) developed a model of the dispersal of a filamentary molecular cloud by an O star forming within the filament. The O star drives an ISF into the filamentary cloud, compressing most of the mass and ionizing only a small fraction of the mass. Applying the equations presented in Section 11 of WP21 to our model filament, we find that after $\sim 1$~Myr the entire filament will have been shocked and condensed into a dense ($\mathrm{n} \sim10^{5}$~cm$^{-3}$) layer/cloud $\sim 0.5$~pc in length. Given the high density of these clouds they could be sites of new star formation activity. The compressed clouds contains 90 percent of the original mass with the other 10 percent of the cloud mass being ionized. Due to the rocket effect caused by the ionized gas streaming away from the clouds, the compressed clouds will develop a recessional velocity relative to the O star of $\sim 2$~km~s$^{-1}$, and after approximately 4~Myr the dense clouds will be 15--20 pc away from the O star when it explodes as a supernova.

\subsection{YSO Identification}
\label{sec:yso}

\subsubsection{Cluster YSO Content}\label{sec:YSOCluster}

UKIDSS J, H, and K photometry was retrieved for cluster members identified using the procedure described in \autoref{sec:cluster}. Poor photometric sources were removed by applying the low completeness--high reliability cuts as described in Appendix A3 of \citet{lucas2008}.

\begin{figure}
    \centering
    \hspace*{-0.75cm}
    \vspace*{-1.20cm}
    \includegraphics[width=1.2\columnwidth]{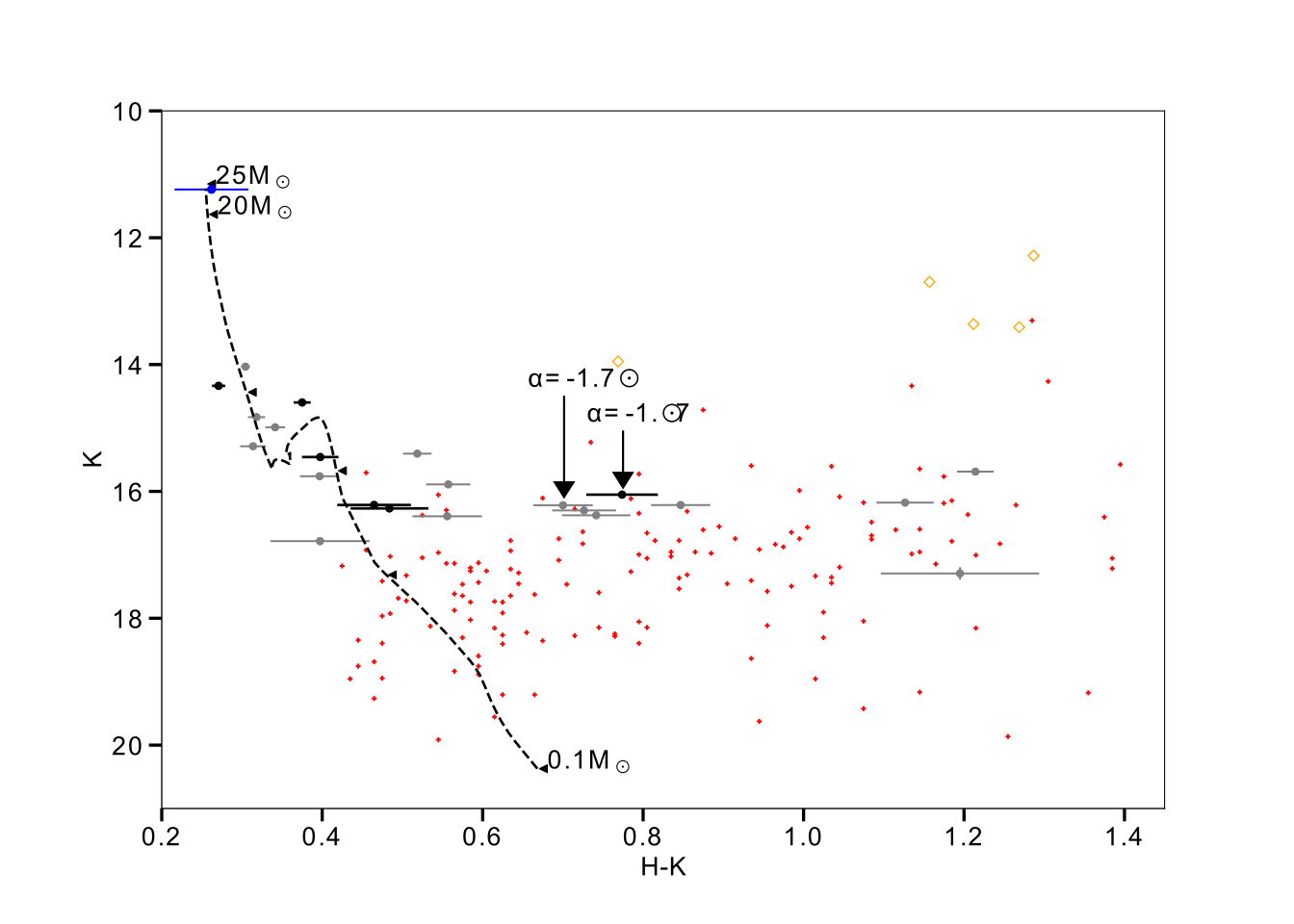}
    \vspace*{0.75cm}
    \caption{CMD of cluster members with various photometric cuts applied. The black dashed line is a 3.65 Myr isochrone. The location of 25, 20 and 0.1 $M_\odot$ stars are labelled, and additional unlabelled tick marks show the location of 5, 2, and 1 $M_\odot$ stars. 2MASS photometry was used for the central ionizing star (plotted as the blue point) as the UKIDSS data were removed by our photometric quality cuts. Grey points represent stars that survived the higher completeness, good photometry cuts from \citet{lucas2008}, while the black points represent stars passing the lower completeness, best photometry cuts. The two IRAC-matched sources are labelled with their respective SED slopes. To illustrate the expected location of YSOs, T Tauri and Herbig AeBe photometry is shown as red dots and gold diamonds respectively (see text for details). 
    }
    \label{fig:ClusterCMD}
\end{figure}

These data were used to construct a color-magnitude diagram (CMD) of the cluster as seen in \autoref{fig:ClusterCMD}. Overlayed on the CMD is an isochrone retrieved from Mesa Isochrones and Stellar Tracks \citep{MIST}, with an age of 3.65 Myr, which matches the estimated age of the \ion{H}{II} region based on the \citet{Schaerer} models for the exciting star. The isochrone was shifted to a distance of 5.99 kpc, then reddened to match the position of the O9 star using $A_V$ = 5.32 mag. This $A_V$ value agrees with  estimates derived from optical photometry reported in \citet{Russeil} and \citet{fos2015}. The isochrone plotted extends from 0.1 to 25 M$_\odot$, and we see that the upper mass limit of the isochrone is in agreement with the 20 $M_\odot$ estimate based on the \citet{Schaerer} OB star evolutionary models. 

In addition to the O star, we see that there are a number of 2 -- 5 M$_\odot$ main sequence stars. There are a number of sources found at larger $H-K$ which are likely young stellar objects (YSOs). To illustrate this, we have included on the diagram representative T Tauri (solar mass) YSOs from \citet{Kenyon}, as well as Herbig Ae/Be (HAeBe, more massive) YSOs from \citet*{The} and \cite{Finkenzeller}. The original photometry of the sources have been shifted to a distance of 5.99 kpc, and the appropriate amount of foreground extinction has been applied. 

The slope ($\alpha$) of the infrared SED, i.e., $\log(F_\lambda)$ versus $\log(\lambda)$ is often used to classify YSOs (e.g., \citealt{kang2017}). Sources with $\alpha > 0.3$ are Type I YSOs (likely pre-main sequence stars surrounded by infalling gas in an envelope and disc structure), sources with $-1.6\leq\alpha\leq-0.3$ are Type II YSOs (likely more evolved stars with a thick disc), and sources with $-3\leq\alpha\leq -1.6$ are Type III YSOs (likely even more evolved stars with only a thin disc of remaining circumstellar material or a bare photosphere). For this analysis we spatially cross-matched (0.5 arcsec match radius) \emph{Spitzer} GLIMPSE360 data with UKIDSS data using the Tool for Operations on Catalogues and Tables ({\sc topcat}; \citealt{TOPCAT}). The addition of the \emph{Spitzer} data is important as it gives us a more accurate estimate of $\alpha$ as the longer wavelength bands are less affected by interstellar reddening. Only two of the potential cluster YSOs (previously identified in the CMD) had matches with \emph{Spitzer} data; one with a match at 3.6~$\mu$m and one with a match at 4.5~$\mu$m. The SEDs were dereddened using the \cite{Flaherty} extinction law and $A_V=5.32$ mag before fitting a slope. The 3.6~$\mu$m matched source (UKIDSSDR11PLUS database source ID 438352769341) is classified as a Type II YSO ($\alpha = -1.37$), and the $4.5~\mu$m matched source (source ID 438352769362) is classified as a Type III YSO ($\alpha = -1.73$). The location of the two sources in the CMD is shown in \autoref{fig:ClusterCMD}.

\begin{figure}
    \centering
    \hspace*{-4.25cm}
    \vspace*{-1.25cm} 
    \includegraphics[width=2\columnwidth]{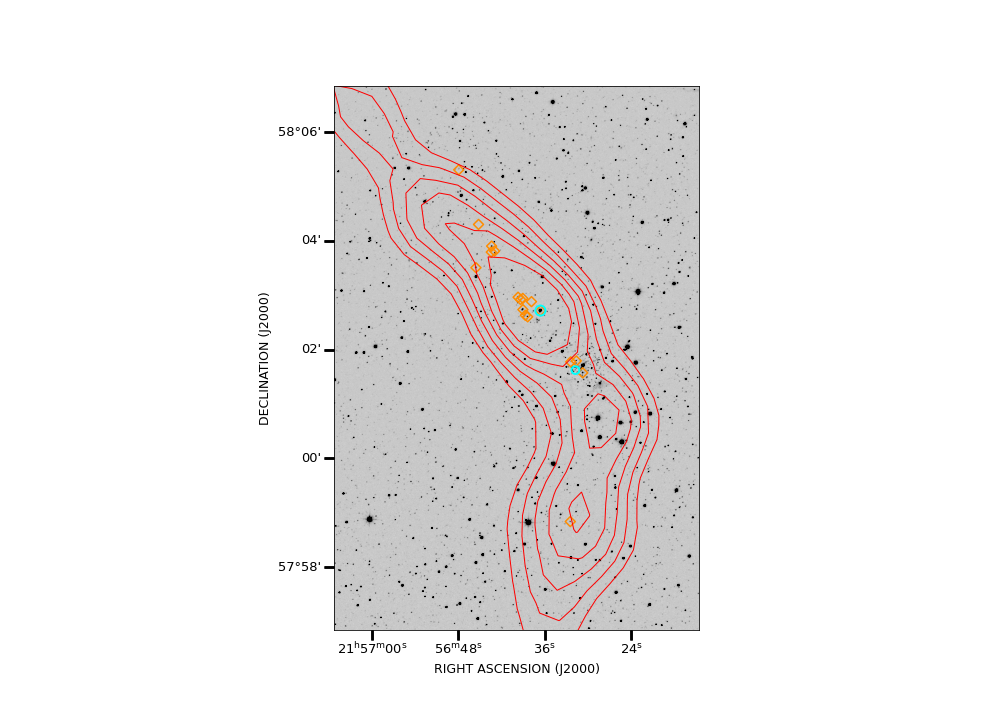}
    \caption{UKIDSS K-band image showing the location of Type II YSOs within the BFS~10 molecular cloud. Red contours (from $1-2.67$~K~km~s$^{-1}$ at 0.33~K~km~s$^{-1}$ intervals) outline the general shape of the molecular cloud. Cyan points represent YSOs identified in this paper, while orange diamonds represent those identified in \citet{win2019}. Two main regions of star formation are evident. One is associated with the \ion{H}{II} region and its embedded rich cluster, and the other is associated with the peak molecular cloud emission in the northern part of the cloud. A few scattered YSOs are also found throughout the molecular cloud with no connection to either of the star-forming regions.
    }
    \label{fig:YSO_Kband}
\end{figure}

\subsubsection{Molecular Cloud YSO Content} \label{sec:YSOCloud}

To determine if star formation activity in the molecular cloud is limited to the embedded cluster, we cross-matched UKIDSS and \emph{Spitzer} photometry of sources found within the extent of the molecular cloud (defined using the 2.67 K~km~s$^{-1}$ contour, see \autoref{fig:YSO_Kband}). This resulted in a sample of 214 sources.

Unfortunately for the molecular cloud, we do not have a good estimate of the extinction to each star as we did for the cluster. Following \cite{lucas2008}, who adopted a locus fitting method devised by \cite{Gutermuth}, we estimate the individual amount of visual extinction using the $H-K$ color excess, denoted as A$_V'$:
\begin{equation}
    A_V' = (H-K-0.2)/0.063 , 
\end{equation}
where the $H-K$ value of 0.2 is chosen as the average intrinsic color of most stars. Using the estimated visual extinction of each star and the \citet{Flaherty} extinction law, we de-reddened each of the remaining sources and constructed their SEDs. Applying the aforementioned SED slope classification to these data revealed no Type I YSOs and one Type II YSO (source ID 438352770201). The remaining 213, are classified, by default, as potential Type III YSOs; however, we cannot differentiate them from unrelated main-sequence stars. 

Visual inspection of mid- and far-IR images of the molecular cloud were done to identify young, heavily-embedded YSOs. No sources are seen in the PACS 70 and 160 $\mu$m images. \emph{Spitzer} 24$\mu$m images revealed a small star-forming region found outside the embedded cluster region boundary and located within the 1 K~km~s$^{-1}$ contour of \autoref{fig:YSO_Kband}. The brightest of these sources is the Type II YSO found in the UKIDSS--\emph{Spitzer} search. Among the remaining other sources, one (source ID 438352770099) was identified in the UKIDSS data set but was removed during the photometric cuts. This source was also noted by \citet{win2019}, and identified as a Type II YSO. 

An additional 18 Type II YSOs within the molecular cloud were identified by \citet{win2019}. These sources are plotted as orange diamonds in \autoref{fig:YSO_Kband}. We did not identify these sources primarily due to our stricter photometric cuts performed on the UKIDSS data, and because \citet{win2019} also included data from the Two-Micron All-Sky Survey \citep[2MASS;][]{skr06}. It is clear that star-formation activity within the molecular cloud is limited to the rich embedded cluster associated with the \ion{H}{ii} region and a sparse grouping of stars associated with an isolated intermediate-mass YSO.

\section{Conclusions} \label{sec:conclusions}

The BFS~10 blister \ion{H}{ii} region is located within a small, filamentary molecular cloud of density and mass similar to an average IRDC. Modeling of the energetics of the region, based on radio and infrared observations, show that the \ion{H}{ii} region is powered by a single O9~V star. Our analysis also shows that the \ion{H}{ii} region has an average geometric covering factor of 0.4, and that the three-dimensional structure of the cloud is truly filamentary as opposed to sheet-like.

Given the filamentary nature of the molecular cloud, BFS~10 is expected to rapidly develop a bipolar morphology in order 10$^5$ yr. It has been suggested that a bipolar \ion{H}{ii} region morphology may arise in cases where massive star-formation has been triggered by colliding molecular clouds, but in this case there is no evidence from the velocity structure of the molecular cloud that the cloud has formed from a cloud-cloud collision. The O-star will eventually compress the filamentary cloud into two compact clouds expanding away from the O-star. These compressed clouds are possible sites of future star formation activity.

A rich embedded cluster within the \ion{H}{II} region was identified using UKIDSS K-band images. The cluster includes 151$\pm$9 stars within a 0.92 pc radius, which is comparable to other known embedded clusters hosting OB-type stars. There are two regions of active star formation found within the molecular cloud, one associated with the embedded star cluster and another sparser group associated with an intermediate-mass YSO. The two regions appear to have evolved independently, and the southern half of the molecular cloud is essentially devoid of star-formation activity. 

\section*{Acknowledgements}
This research has made use of the NASA/IPAC Infrared Science Archive, which is funded by the National Aeronautics and Space Administration and operated by the California Institute of Technology. This research used the facilities of the Canadian Astronomy Data Centre operated by the National Research Council of Canada with the support of the Canadian Space Agency.

\section*{Data Availability}

The majority of the data underlying this article were accessed from the CADC (https://www.cadc-ccda.hia-iha.nrc-cnrc.gc.ca/), IRSA (http:irsa.ipac.caltech.edu/), and the WFCAM Science Archive (http://wsa.roe.ac.uk/). The derived data generated in this research will be shared on reasonable request to the corresponding author. The molecular line data underlying this article will also be shared on reasonable request to the corresponding author.

\bibliographystyle{mnras}
\bibliography{BFS10}

\bsp	
\label{lastpage}
\end{document}